\documentclass[twocolumn,preprintnumbers,superscriptaddress,nofootinbib]{revtex4}
\usepackage{amsmath,amssymb}
\usepackage{graphicx}
\usepackage{epsfig,color}
\usepackage{bm}
\usepackage{verbatim}

\begin{document}

\title{The baryo-quarkonium picture for hidden-charm and bottom pentaquarks and LHCb $P_{\rm c}(4380)$ and $P_{\rm c}(4450)$ states}

\author{J. Ferretti}
\affiliation{CAS Key Laboratory of Theoretical Physics, Institute of Theoretical Physics, Chinese Academy of Sciences, Beijing 100190, China}
\affiliation{Center for Theoretical Physics, Sloane Physics Laboratory, Yale University, New Haven, Connecticut 06520-8120, USA}
\author{E. Santopinto}
\affiliation{INFN, Sezione di Genova, Via Dodecaneso 33, 16146 Genova, Italy}
\author{M. Naeem Anwar}
\affiliation{CAS Key Laboratory of Theoretical Physics, Institute of Theoretical Physics, Chinese Academy of Sciences, Beijing 100190, China}
\affiliation{University of Chinese Academy of Sciences, Beijing 100049, China}
\author{M. A. Bedolla}
\affiliation{INFN, Sezione di Genova, Via Dodecaneso 33, 16146 Genova, Italy}
\affiliation{Instituto de F\'isica y Matem\'aticas, Universidad Michoacana de San Nicol\'as de Hidalgo, Edificio C-3, Ciudad Universitaria, Morelia, Michoac\'an 58040, M\'exico}

\begin{abstract}
We study baryo-charmonium [$\eta_{\rm c}$- and $J/\psi$-$N^*$, $\eta_{\rm c}(2S)$-, $\psi(2S)$- and $\chi_{\rm c}(1P)$-$N$] and baryo-bottomonium [$\eta_{\rm b}(2S)$-, $\Upsilon(2S)$- and $\chi_{\rm b}(1P)$-$N$] bound states, where $N$ is the nucleon and $N^*$ a nucleon resonance.
In the baryo-quarkonium model, the five $qqq Q \bar Q$ quarks are arranged in terms of a heavy quarkonium core, $Q \bar Q$, embedded in light baryonic matter, $qqq$, with $q = u$ or $d$.
The interaction between the $Q \bar Q$ core and the light baryon can be written in terms of the QCD multipole expansion.
The spectrum of baryo-charmonium states is calculated and the results compared with the existing experimental data. In particular, we can interpret the recently discovered $P_{\rm c}(4380)$ and $P_{\rm c}(4450)$ pentaquarks as $\psi(2S)$-$N$ and $\chi_{\rm c2}(1P)$-$N$ bound states, respectively.
We observe that in the baryo-bottomonium sector the binding energies are, on average, slightly larger than those of baryo-charmonia. Because of this, the hidden-bottom pentaquarks are more likely to form than their hidden-charm counterparts. We thus suggest the experimentalists to look for five-quark states in the hidden-bottom sector in the $10.4-10.9$ GeV energy region.
\end{abstract}

\maketitle

\section{Introduction}
Recently, LHCb reported the observation of two new resonances, $P_c^+(4380)$ and $P_c^+(4450)$, in $\Lambda_{\rm b} \rightarrow J/\psi \Lambda^*$ and $\Lambda_{\rm b} \rightarrow P_{\rm c}^+ K^- \rightarrow (J/\psi p) K^-$ decays \cite{Aaij:2015tga}.
Their quark structure is $\left| P_{\rm c}^+ \right\rangle = \left| uud c \bar c \right\rangle$, whence the name pentaquarks.
The pentaquarks were introduced in the LHCb analysis of $\Lambda_{\rm b}$ decays to improve the fit upon the experimental data, because the use of known $\Lambda^*$ states alone was not sufficient to get a satisfactory description of the $J/\psi p$ spectrum \cite{Nakamura:2010zzi}.
The pentaquarks masses, resulting from the LHCb best fit, are $M_{P_{\rm c}^+(4380)} = 4380\pm8\pm29$ and $M_{P_{\rm c}^+(4450)} = 4449.8\pm1.7\pm2.5$ MeV, with widths $\Gamma_{P_c^+(4380)} = 205\pm18\pm86$ and $\Gamma_{P_c^+(4450)} = 39\pm5\pm19$ MeV. The preferred $J^P$ quantum numbers are $(\frac{3}{2}^-,\frac{5}{2}^+)$, even if $(\frac{3}{2}^+,\frac{5}{2}^-)$ and $(\frac{5}{2}^+,\frac{3}{2}^-)$ are also acceptable solutions \cite{Aaij:2015tga}; indeed, all the preferred fits from LHCb require pentaquarks with opposite parities.

From a theoretical point of view, there are a few possible interpretations for a five-quark bound state, including: I) Baryon-meson molecules \cite{bszou,Yang:2011wz,Feijoo:2015cca,He:2015cea,Karliner:2015ina,Chen:2015loa,Chen:2015moa,Azizi:2016dhy,Yamaguchi:2016ote,Ortega:2016syt,Dong:2016dkh}, such as $\Sigma_{\rm c}^+ \bar D^{*0}$, the $P_{\rm c}^+(4450)$ lying 10 MeV below the $\Sigma_{\rm c}^+ \bar D^{*0}$ threshold. Other molecular model assignments, like $\Sigma_{\rm c}^* \bar D$, are also possible; II) Diquark-diquark-antiquark states \cite{Maiani:2015vwa,Lebed:2015tna,Wang:2015epa,Li:2015gta}, made up of a charm antiquark, $\bar c$, a heavy-light diquark, $[cq]$, and a light-light one, $[qq]$, where $q = u$ or $d$; III) Baryo-charmonium systems \cite{Eides:2015dtr,Perevalova:2016dln,Alberti:2016dru}, like $\psi(2S)$-$N$, bounded by gluon-exchange forces; IV) The result of kinematical or threshold-rescattering effects \cite{Liu:2015fea,Meissner:2015mza,Guo:2015umn}, like in the case of $\chi_{\rm c1}p$ resonances or anomalous threshold singularities; V) The bound state of open-color configurations \cite{Mironov:2015ica}; VI) The bound states of a soliton and two pseudoscalar mesons, $D$ and $\bar D$ \cite{Scoccola:2015nia}; VII) Compact five-quark systems \cite{Yang:2018oqd,Ortiz-Pacheco:2018ccl,Santopinto:2016pkp}.
For a review, see Refs. \cite{Chen:2016qju,Ali:2017jda,Olsen:2017bmm,Guo:2017jvc}.
In this work, we discuss the baryo-charmonium (baryo-quarkonium) one.

The hypothesis of charmonium-nuclei bound states dates back to the early nineties. At that time, it was shown that QCD van der Waals-type interactions, due to multiple gluon-exchange, may provide a strong enough binding to produce charmonium-nuclei bound states if $A \gtrsim 4$ \cite{Brodsky:1989jd,Luke:1992tm,Kaidalov:1992hd,Sibirtsev:2005ex}, where $A$ is the atomic mass number.
On studying the charmonium-nucleon systems, the interaction, though attractive [$\mathcal O(10)$ MeV], is too weak to produce a $c \bar c$-$N$ bound state.
Notwithstanding, it is still unclear if a similar interaction may give rise to $c \bar c$-$qqq$ bound states if the nucleon is replaced by its radial or orbital excitations, or the charmonium ground-state by its radial excitations.
These possibilities are worth to be investigated in the baryo-charmonium picture.

By analogy with four-quark hadro-charmonia \cite{Dubynskiy:2008mq,Guo:2008zg,Voloshin:2013dpa,Wang:2013kra,Brambilla:2015rqa,Panteleeva:2018ijz,Ferretti:2018kzy}, namely $c \bar c$-$q \bar q$ states, the baryo-charmonium is a pentaquark configuration, where a compact $c \bar c$ state, $\psi$, is embedded in light baryonic matter, $\mathcal B$ \cite{Eides:2015dtr,Perevalova:2016dln,Alberti:2016dru}.
The interaction between the two components, $\psi$ and $\mathcal B$, takes place via a QCD analog of the van der Waals force of molecular physics.
It can be written in terms of the multipole expansion in QCD \cite{QCDME}, with the leading term being the $E1$ interaction with chromo-electric field ${\bf E}^a$.

In the present manuscript, we use the baryo-charmonium model to discuss the possible emergence of $\eta_{\rm c}$- and $J/\psi$-$N^*$, $\eta_{\rm c}(2S)$-, $\psi(2S)$- and $\chi_{\rm c}(1P)$-$N$ bound states, where $N$ is the nucleon and $N^*$ a nucleon resonance.
The energies of baryo-charmonia are computed by solving the Schr\"odinger equation for the baryo-charmonium potential \cite{Dubynskiy:2008mq,Ferretti:2018kzy}. This is approximated as a finite well whose width and size can be expressed as a function of the $N$ ($N^*$) radius and the charmonium chromo-electric polarizability, $\alpha_{\psi \psi}$.
The baryo-charmonium masses and quantum numbers are compared with the existing experimental data and some tentative assignments are discussed; in particular, we can interpret the recently discovered $P_{\rm c}(4380)$ and $P_{\rm c}(4450)$ pentaquarks as $\psi(2S)$-$N$ and $\chi_{\rm c2}(1P)$-$N$ bound states, respectively.

Furthermore, we extend the previous calculations to the bottom sector and calculate the spectrum of bottomonium-$N$ bound states.
Our results are compatible with the emergence of $2S$ and $1P$ bottomonium-nucleon bound states, with binding energies of the order of a few hundreds of MeV.
We observe that in the baryo-bottomonium sector the binding energies are, on average, slightly larger than those of baryo-charmonia. Because of this, the hidden-bottom pentaquarks are more likely to form than their hidden-charm counterparts. We thus suggest the experimentalists to look for five-quark states in the hidden-bottom sector in the $10.4-10.9$ GeV energy region.

\section{Baryo-quarkonium Hamiltonian}
\label{Baryo-charmonium Hamiltonian}
The baryo-quarkonium is a particular pentaquark configuration, where five quarks are arranged in terms of a compact $Q \bar Q$ state embedded in light baryonic matter.
The interaction between the quarkonium core, $\mathcal Q$, and the gluonic field inside the light-baryon, $\mathcal B$, can be written in terms of the QCD multipole expansion \cite{QCDME,Anwar:2016mxo}.
In particular, one considers as leading term the $E1$ interaction with chromo-electric field ${\bf E}$ \cite{Dubynskiy:2008mq,Kaidalov:1992hd},
\begin{equation}
	\label{eqn:Heff}
	H_{\rm eff} = - \frac{1}{2} \alpha_{ij} {\bf E}_i \cdot {\bf E}_j  \mbox{ },
\end{equation}
$\alpha_{ij}$ being the quarkonium chromo-electric polarizability.
In order to calculate the baryo-quarkonium masses, one has to compute the expectation value of Eq. (\ref{eqn:Heff}) on $\left| \mathcal Q \mathcal B \right\rangle$ states.

The chromo-electric field matrix elements can be calculated in terms of the QCD energy-momentum tensor, $\theta_{\mu}^{\mu} \approx \frac{9}{16 \pi^2} {\bf E}^2$ \cite{Voloshin:1980zf}.
Its expectation value on a nonrelativistic normalized $\left| \mathcal B \right\rangle$ at rest gives the mass of this state \cite{Dubynskiy:2008mq},
\begin{equation}
  \label{eqn:thetamumu}
	\left\langle \mathcal B \right| \theta_\mu^\mu ({\bf q} = 0) \left| \mathcal B \right\rangle \simeq M_{\mathcal B}  \mbox{ }.
\end{equation}
The baryo-quarkonium effective potential, $V_{\rm bq}$, describing the coupling between $\mathcal Q$ and $\mathcal B$, can be approximated by a finite well \cite{Dubynskiy:2008mq,Ferretti:2018kzy}
\begin{equation}
	\label{eqn:Vd3r}
	\int_0^{R_{\mathcal B}} d^3r \mbox{ } V_{\rm bq} \approx - \frac{8 \pi^2}{9} \mbox{ } \alpha_{\mathcal Q \mathcal Q} M_{\mathcal B}  \mbox{ },
\end{equation}
where $R_{\mathcal B}$ is the radius of $\mathcal B$ \cite{Nakamura:2010zzi,Santopinto:2014opa} and $\alpha_{\mathcal Q \mathcal Q}$ the quarkonium diagonal chromo-electric polarizability.
Thus, we have:
\begin{equation}
	\label{eqn:Vbc}
	V_{\rm bq}(r) = \left\{ \begin{array}{ccc} -\frac{2\pi\alpha_{\mathcal Q \mathcal Q}M_{\mathcal B}}{3R_{\mathcal B}^3} & \mbox{for} & r < R_{\mathcal B} \\
	                         0   & \mbox{for} & r > R_{\mathcal B}
		                \end{array}  \right.  \mbox{ }.
\end{equation}
The kinetic energy term is 
\begin{equation}
	T_{\rm bq} = \frac{k^2}{2 \mu} \mbox{ },
\end{equation}
where ${\bf k}$ is the relative momentum (with conjugate coordinate ${\bf r}$) between $\mathcal Q$ and $\mathcal B$, and $\mu$ the reduced mass of the $\mathcal Q \mathcal B$ system.
Finally, the total baryo-quarkonium Hamiltonian is:
\begin{equation}
	\label{eqn:Hhc}
	H_{\rm bq} = M_{\mathcal Q} + M_{\mathcal B} + V_{\rm bq}(r) + T_{\rm bq}  \mbox{ }.
\end{equation}

The baryo-quarkonium quantum numbers are obtained by combining those of the quarkonium core, $\mathcal Q$, and light baryon, $\mathcal B$, as
\begin{equation}
\label{quantNumbers}
	\left| \Phi_{\rm bq} \right\rangle = \left| (L_\mathcal Q, S_{\mathcal Q}) J_{\mathcal Q}; (L_{\mathcal B}, S_{\mathcal B}) J_{\mathcal B}; (J_{\rm bq}, \ell_{\rm bq}) J_{\rm tot}^P \right\rangle  \mbox{ },
\end{equation}
where ${\bf J}_{\rm bq} = {\bf J}_{\mathcal Q} + {\bf J}_{\mathcal B}$, the baryo-quarkonium parity is $P = (-1)^{\ell_{\rm bq}}$ $P_{\mathcal Q} P_{\mathcal B}$, and $\ell_{\rm bq}$ is the relative angular momentum between $\mathcal Q$ and $\mathcal B$. 
From now on, unless we indicate an explicit value, we will assume $\ell_{\rm bq} = 0$.

\section{Chromo-electric polarizability}
\label{Chromo-electric polarizability}
In this section, we depict two different procedures for the quarkonium diagonal chromo-electric polarizabilities. 

\subsection{Chromo-electric polarizabilities of charmonia as pure Coulombic systems}
\label{Chromo-electric polarizabilities of cc}
There are several possible approaches for the quarkonium diagonal chromo-electric polarizability.
One possibility is to calculate it by considering quarkonia as pure Coulombic systems.
The perturbative result in the framework of the $1/N_{\rm c}$ expansion is \cite{Peskin:1979va}
\begin{equation}
	\label{alphacoulomb}
	\alpha_{\psi \psi}(nS)=\frac{16\pi n^2 c_n a_0^3}{3g_{\rm c}^2 N_{\rm c}^2} \mbox{ }.
\end{equation}
Here, $n$ is the radial quantum number; $c_1=\frac{7}{4}$ and $c_2=\frac{251}{8}$; $N_{\rm c} = 3$ is the number of colors; $g_{\rm c} = \sqrt{4\pi \alpha_{\rm s}} \simeq 2.5$, and $\alpha_{\rm s}$ is the QCD running coupling constant at the charm quark mass-scale; finally,
\begin{equation}
	\label{eqn:a0-Bohr}
	a_0 = \frac{2}{m_{\rm c} C_{\rm F} \alpha_{\rm s}}
\end{equation}
is the Bohr radius of nonrelativistic charmonium \cite{Brambilla:2015rqa}, with color factor $C_{\rm F} = \frac{N_{\rm c}^2 - 1}{2 N_{\rm c}}$, and $m_{\rm c} = 1.5$ GeV is the  charm quark mass.
A nonperturbative calculation of the chromo-electric polarizability was carried out in Refs. \cite{Leutwyler:1980tn,Voloshin:1979uv}. The result is
\begin{equation}
	\label{alphacoulomb-exact}
	\alpha_{\psi \psi}(n\ell) = \frac{2 n^6 \epsilon_{n \ell}}{m_{\rm c}^3 \mbox{ } \beta^4} \mbox{ },
\end{equation}
where $\ell$ is the orbital angular momentum, $\beta = \frac{4}{3} \alpha_{\rm s}$ and $\epsilon_{n \ell}$'s are numerical coefficients, with $\epsilon_{1 0} = 1.468$.
The use of Eqs. (\ref{alphacoulomb}, \ref{eqn:a0-Bohr}) or (\ref{alphacoulomb-exact}) provides the same result. One obtains \cite{Ferretti:2018kzy}
\label{eqn:alpha-Coul}
\begin{equation}
	\alpha_{\psi \psi}^{\rm Coul}(1S) \simeq 4.1 \mbox{ GeV}^{-3}
\end{equation}	
and
\begin{equation}
	\alpha_{\psi \psi}^{\rm Coul}(2S) \simeq 296 \mbox{ GeV}^{-3}  \mbox{ }.
\end{equation}

%%%%%%%%%%%%%%%%%%%%%%%%%%
\begin{table}[htbp]
\centering
\begin{tabular}{ccccc}
\hline
\hline
Source & $a_{N \psi}$ & $\alpha_{\psi \psi}^{\rm scatt}(1S)$ & $\alpha_{\psi \psi}^{\rm scatt}(2S)$ & $\alpha_{\psi \psi}^{\rm scatt}(1P)$ \\
            &               [fm] &                   [GeV$^{-3}$] &  [GeV$^{-3}$] &  [GeV$^{-3}$] \\
\hline
\cite{Kaidalov:1992hd,Gryniuk:2016mpk} & $\approx -0.05$ & 0.25 & 18 & 11 \\
\cite{Sibirtsev:2005ex,Voloshin:2007dx} & $-0.37$ & 2.0 & 143 & 90 \\
\cite{Yokokawa:2006td} & $-0.70\pm0.66$ & 3.8 & 273 & 172 \\
\cite{Brodsky:1997gh} & $-0.24$ & 1.3 & 93 & 59 \\
\cite{Hayashigaki:1998ey} & $-0.10 \pm 0.02$ & 0.54 & 39 & 25 \\
\cite{Kawanai:2010ev} & $\approx 0.3$ & 1.6 & 115 & 72 \\
\hline
\hline
\end{tabular}
\caption{The chromo-electric polarizabilities, $\alpha_{\psi \psi}^{\rm scatt}$, are fitted to charmonium-nucleon scattering lengths, $a_{N \psi}$, via Eq. (\ref{eqn:Npsi-scatt-length}).}
\label{tab:alpha-scatt}
\end{table}
%%%%%%%%%%%%%%%%%%%%%%%%%%

\subsection{Chromo-electric polarizabilities from charmonium-nucleon scattering lengths}
The second approach to extract charmonium diagonal chromo-electric polarizabilities is to fit them to results of charmonium-nucleon scattering lengths.
The latter can be written as \cite[Eq. (104)]{Krein:2017usp}
\begin{equation}
	\label{eqn:Npsi-scatt-length}
	a_{N \psi} \approx - \frac{4 \pi M_N}{9} \mu \alpha_{\psi \psi}  \mbox{ },
\end{equation}
where $M_N$ is the nucleon mass and $\mu$ the charmonium-nucleon reduced mass. The results are shown in Table \ref{tab:alpha-scatt}. It is interesting to observe that the calculated values of $\alpha_{\psi \psi}(1S)$ span a wide interval, $\alpha_{\psi \psi} \in [0.25 - 3.8]$ GeV$^{-3}$.

In particular, a global fit to both differential and total cross sections from available data on $J/\psi p$ scattering provides a value $a_{p J/\psi} = -0.046\pm0.005$ fm \cite{Gryniuk:2016mpk}, which is consistent with the value $a_{N \eta_{\rm c}} = -0.05$ fm from Ref. \cite{Kaidalov:1992hd}. The corresponding value of the binding energy of $J/\psi$ in nuclear matter is $\approx 3$ MeV, which is close to the deuteron binding energy.
On the other hand, the binding energy of $J/\psi$ in nuclear matter was found to be 21 MeV for $\alpha_{\psi \psi}(1S) = 2$ GeV$^{-3}$ \cite{Sibirtsev:2005ex,Voloshin:2007dx}, corresponding to a scattering length of $a_{N J/\psi} = -0.37$ fm.
An even larger value for the charmonium-nucleon scattering length was obtained by means of quenched lattice QCD calculations, $a_{N \psi} \approx -0.7$ fm \cite{Yokokawa:2006td}.

Finally, we extract the values of the chromo-electric polarizabilities of $2S$ and $1P$ charmonia.
That of $2S$ states can be estimated as four times the ratio between $c_2=\frac{251}{8}$ and $c_1=\frac{7}{4}$. One gets
\begin{equation}
	\alpha_{\psi \psi}(2S) = \frac{502}{7} \mbox{ } \alpha_{\psi \psi}(1S)  \mbox{ }.
\end{equation}
The chromo-electric polarizability of $1P$ charmonia can be estimated by means of Eq. (\ref{alphacoulomb-exact}) and Ref.~\cite[Table 1]{Leutwyler:1980tn}. This means
\begin{equation}
   \label{alpha1P}
	\alpha_{\psi \psi}(1P) = \frac{\epsilon_{21}}{\epsilon_{20}} \mbox{ } \alpha_{\psi \psi}(2S)  \mbox{ },
\end{equation}
where $\epsilon_{20} = 1.585$ and $\epsilon_{21} = 0.998$.

It is still not possible to fit $\alpha_{\psi \psi}(n\ell)$ to the experimental data. 
So, as previously discussed, $\alpha_{\psi \psi}(n\ell)$'s have to be estimated phenomenologically. This could be one of the main sources of theoretical uncertainty on our results. 

\subsection{Chromo-electric polarizabilities of bottomonia}
\label{Chromo-electric polarizability of bb}
Finally, we calculate bottomonium diagonal chromo-electric polarizabilities by considering them as pure Coulombic systems. See Eqs. (\ref{alphacoulomb}) and (\ref{eqn:a0-Bohr}), where we substitute the charm-quark mass, $m_{\rm c}$, with the bottom one, $m_{\rm b} = 5.0$ GeV, and evaluate $\alpha_s$ at the $m_{\rm b}$ mass-scale. We get
\begin{equation}
	\alpha_{\Upsilon \Upsilon}^{\rm max}(1S) \simeq 0.47 \mbox{ GeV}^{-3}
\end{equation}	
and
\begin{equation}
	\label{eqn:alpha-2S-max}
	\alpha_{\Upsilon \Upsilon}^{\rm max}(2S) \simeq 33 \mbox{ GeV}^{-3}  \mbox{ }.
\end{equation}
Similar values are obtained by using the nonperturbative results of Refs. \cite{Leutwyler:1980tn,Voloshin:1979uv}.
On the contrary, if we define the bottomonium Bohr radius as \cite{Eides:2015dtr,Perevalova:2016dln}
\begin{equation}
	\label{eqn:a0-Bohr2}
	a_0 = \frac{16 \pi}{g_{\rm b}^2 N_{\rm c} m_{\rm b}}  \mbox{ },
\end{equation}
we get
\begin{equation}
	\alpha_{\Upsilon \Upsilon}^{\rm min}(1S) \simeq 0.33 \mbox{ GeV}^{-3}
\end{equation}	
and
\begin{equation}
	\label{eqn:alpha-2S-min}
	\alpha_{\Upsilon \Upsilon}^{\rm min}(2S) \simeq 23 \mbox{ GeV}^{-3}  \mbox{ }.
\end{equation}
The chromo-electric polarizabilities of $1P$ bottomonia can be estimated using Eq.~(\ref{alpha1P}); we get
\begin{equation}
	\label{eqn:alpha-1P-max}
	\alpha_{\Upsilon \Upsilon}^{\rm max}(1P) \simeq 21 \mbox{ GeV}^{-3}
\end{equation}	
and
\begin{equation}
	\label{eqn:alpha-1P-min}
	\alpha_{\Upsilon \Upsilon}^{\rm min}(1P) \simeq 14 \mbox{ GeV}^{-3}  \mbox{ }.
\end{equation}

%%%%%%%%%%%%%%%%%%%%%%%%%%
\begin{table*}[htbp]
\centering
\begin{tabular}{ccccccccccc}
\hline
\hline
Composition & & $\alpha_{\psi \psi}(n\ell)$ [GeV$^{-3}$] & & $J_{\rm tot}^P$ & & Binding [MeV] & & Mass [MeV] & & Assignment \\
\hline
$\eta_{\rm c} \otimes N(1440)$ & & 4.1 & & $\frac{1}{2}^-$  & & $-16$ & & 4397 & & -- \\
$\eta_{\rm c} \otimes N(1520)$ & & 4.1 & & $\frac{3}{2}^+$ & & $-22$ & & 4476 & & -- \\
$\eta_{\rm c} \otimes N(1535)$ & & 4.1 & & $\frac{1}{2}^+$ & & $-23$ & & 4495 & & -- \\
$J/\psi \otimes N(1440)$ & & 4.1 & & $\frac{1}{2}^-$ or $\frac{3}{2}^-$ & & $-17$ & & 4510 & & -- \\
$J/\psi \otimes N(1520)$ & & 4.1 & & $\frac{1}{2}^+$, $\frac{3}{2}^+$ or $\frac{5}{2}^+$ & & $-23$ & & 4589 & & -- \\
$J/\psi \otimes N(1535)$ & & 4.1 & & $\frac{1}{2}^+$, $\frac{3}{2}^+$ or $\frac{5}{2}^+$ & & $-24$ & & 4608 & & -- \\
\hline
$\chi_{\rm c0}(1P) \otimes N$ & & 11 & & $\frac{1}{2}^+$ & & $-83$ & & 4274 & & -- \\
$\eta_{\rm c}(2S) \otimes N$ & & 18 & & $\frac{1}{2}^-$ & & $-213$ & & 4365 & & -- \\
$\chi_{\rm c1}(1P) \otimes N$ & & 11 & & $\frac{1}{2}^+$ or $\frac{3}{2}^+$ & & $-84$ & & 4366 & & -- \\
$h_{\rm c}(1P) \otimes N$ & & 11 & & $\frac{1}{2}^+$ or $\frac{3}{2}^+$ & & $-84$ & & 4381 & & -- \\
$\chi_{\rm c2}(1P) \otimes N$ & & 11 & & $\frac{1}{2}^+$, $\frac{3}{2}^+$ or $\frac{5}{2}^+$ & & $-84$ & & 4411 & & $P_{\rm c}(4450)$ \\
$\psi(2S) \otimes N$ & & 18 & & $\frac{1}{2}^-$ or $\frac{3}{2}^-$ & & $-213$ & & 4412 & & $P_{\rm c}(4380)$ \\
$\left[\eta_{\rm c}(2S) \otimes N\right]_{\ell_{\rm bc} = 1}$ & & 18 & & $\frac{1}{2}^+$ or $\frac{3}{2}^+$ & & $-37$ & & 4542 & & -- \\
$\left[\psi(2S) \otimes N\right]_{\ell_{\rm bc} = 1}$ & & 18 & & $\frac{1}{2}^+$, $\frac{3}{2}^+$ or $\frac{5}{2}^+$ & & $-37$ & & 4588 & & -- \\
\hline
\hline
\end{tabular}
\caption{Baryo-charmonium model predictions (fourth and fifth columns), calculated by solving the Schr\"odinger equation (\ref{eqn:Hhc}) with the chromo-electric polarizabilities $\alpha_{\psi \psi}^{\rm Coul}(1S)$ (upper part of the table) or $\alpha_{\psi \psi}^{\rm scatt}(2S)$ and $\alpha_{\psi \psi}^{\rm scatt}(1P)$ (lower part).}
\label{tab:baryo-charmonium-spectrum}
\end{table*}
%%%%%%%%%%%%%%%%%%%%%%%%%%

\section{Baryo-charmonia and the $P_{\rm c}(4380)$ and $P_{\rm c}(4450)$ pentaquarks}
In this section, we give results for the binding energies of charmonium-$N$ and $N^*$ bound states. The previous observables are computed by using the values of the charmonium chromo-electric polarizabilities from Sec. \ref{Chromo-electric polarizability}.

\begin{figure}[htbp]
\centering
\includegraphics[width=8cm]{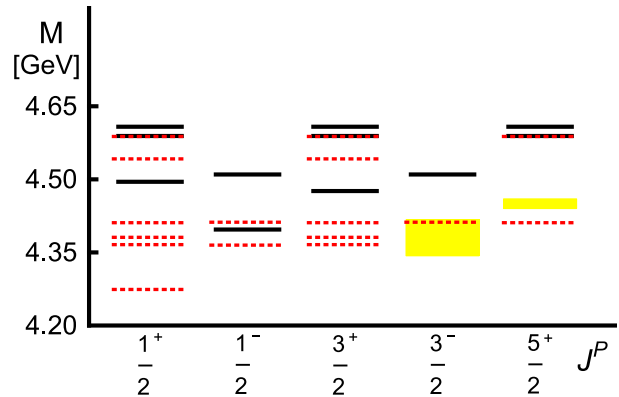}
\caption{Baryo-charmonium spectrum of $\eta_{\rm c}(2S)$-, $\psi(2S)$-, and $\chi_{\rm c}(1P)$-$N$ bound states (dotted lines), calculated with $\alpha_{\psi \psi}^{\rm scatt}(2S) = 18$ GeV$^{-3}$ and $\alpha_{\psi \psi}^{\rm scatt}(1P) = 11$ GeV$^{-3}$, and  $\eta_{\rm c}$- and $J/\psi$-$N^*$ bound states (black lines), calculated with $\alpha_{\psi \psi}^{\rm Coul}(1S) = 4.1$ GeV$^{-3}$. The theoretical results are compared to the experimental masses of $P_{\rm c}(4380)$ and $P_{\rm c}(4450)$ pentaquarks (boxes) \cite{Aaij:2015tga}.}
\label{fig:bc-spectrum}
\end{figure}

The spectrum of $\eta_{\rm c}$- and $J/\psi$-$N^*$, $\eta_{\rm c}(2S)$-, $\psi(2S)$- and $\chi_{\rm c}(1P)$-$N$ bound states is calculated in the baryo-charmonium picture by solving the eigenvalue problem of Eq. (\ref{eqn:Hhc}). See Table \ref{tab:baryo-charmonium-spectrum} and Fig. \ref{fig:bc-spectrum}.
The time-independent Schr\"odinger equation is solved numerically by means of a finite differences algorithm \cite[Vol. 3, Sec. 16-6]{Feynman-Lectures}.
The results strongly depend on the values of charmonium diagonal chromo-electric polarizabilities, $\alpha_{\psi \psi}(n\ell)$.
These values are not defined unambiguously, but span a wide interval.
Up to now, $\alpha_{\psi \psi}(n\ell)$'s cannot be fitted to the experimental data; they have to be estimated phenomenologically. Because of this, they represent one of the main sources of theoretical uncertainty on our results.
We have thus decided to present two sets of results for the baryo-charmonium spectrum.
%%%%%%%%%%%%%%%%%%%%%%%%%%%%%%%%%%%%%%%%%%%%%%%%%%
%%%%%%%%%%%%%%%%%%%%%%%%%%%%%%%%%%%%%%%%%%%%%%%%%%

In the first case, we use the values $\alpha_{\psi \psi}^{\rm scatt}(1S) = 0.25$ GeV$^{-3}$, $\alpha_{\psi \psi}^{\rm scatt}(2S) = 18$ GeV$^{-3}$ and $\alpha_{\psi \psi}^{\rm scatt}(1P) = 11$ GeV$^{-3}$, corresponding to a charmonium-nucleon scattering length $a_{N J/\psi} \simeq a_{N \eta_{\rm c}} \simeq -0.05$ fm \cite{Gryniuk:2016mpk,Kaidalov:1992hd}. These values of $\alpha_{\psi \psi}(n\ell)$ are of the same order of magnitude as those of \cite[Eq. (4)]{Eides:2015dtr}.
Baryo-charmonium bound states will give rise due to the interaction between $N$ and $2S$ and $1P$ charmonia.

In the second case, we use $\alpha_{\psi \psi}^{\rm Coul}(1S) = 4.1$ GeV$^{-3}$ and $\alpha_{\psi \psi}^{\rm Coul}(2S) = 296$ GeV$^{-3}$.
These values are extracted by considering charmonia as pure Coulombic systems \cite{Peskin:1979va,Leutwyler:1980tn,Voloshin:1979uv}.
Similar values of $\alpha_{\psi \psi}^{\rm Coul}(1S)$ are obtained by fitting the $1S$ chromo-electric polarizability to the quenched lattice QCD result for the charmonium-nucleon scattering length of Ref. \cite{Yokokawa:2006td}.
Thus, the baryo-charmonium bound states will give rise due to the interaction between $N^*$ and $1S$ charmonia, $\alpha_{\psi \psi}^{\rm Coul}(2S)$ being too large \cite{Ferretti:2018kzy}.

It is worth noticing that we are able to make at least a clear assignment in the $\alpha_{\psi \psi}^{\rm scatt}(2S)$ case. In particular, the $P_{\rm c}(4380)$ pentaquark is interpreted as a $\psi(2S) \otimes N$ bound state with $J^P = \frac{3}{2}^-$ quantum numbers.
Besides, we can also speculate on assigning the $P_{\rm c}(4450)$ to a $\chi_{\rm c2}(1P) \otimes N$ baryo-charmonium state although, in this second case, the theoretical prediction for the mass falls outside the experimental mass interval.

However, our predictions do not agree with the baryo-charmonium results of Refs. \cite{Eides:2015dtr,Perevalova:2016dln}, where the $P_{\rm c}(4450)$ pentaquark is interpreted as a $\psi(2S) \otimes N$ bound state. This difference could be related to the different choices on $\alpha_{\psi \psi}(2S)$: our values are calculated with 18 GeV$^{-3}$ and theirs with 12 GeV$^{-3}$.

\section{Baryo-bottomonium pentaquark states}
Below, we calculate the spectrum of $\eta_{\rm b}(2S)$-, $\Upsilon(2S)$- and $\chi_{\rm b}(1P)$-$N$ bound states in the baryo-bottomonium picture by solving the eigenvalue problem of Eq. (\ref{eqn:Hhc}). The results are enlisted in Table~\ref{tab:baryo-bottomonium-spectrum}. 
The baryo-bottomonium quantum numbers, shown in the third column of Table \ref{tab:baryo-bottomonium-spectrum}, are obtained by means of the prescription of Eq. (\ref{quantNumbers}).
%%%%%%%%%%%%%%%%%%%%%%%%%%%%%%%%%%%%%%%%%%%%%%%%%%
\begin{table*}[htbp]
\centering
\begin{tabular}{ccccccccc}
\hline
\hline
Composition & & $\alpha_{\Upsilon \Upsilon}(n\ell)$ [GeV$^{-3}$] & & $J_{\rm tot}^P$ & & Binding [MeV] & & Mass [MeV] \\
\hline
$\eta_{\rm b}(2S) \otimes N$ & & 23 & & $\frac{1}{2}^-$ & & $-333$ & & 10605  \\
$\Upsilon(2S) \otimes N$ & & 23 & & $\frac{1}{2}^-$ or $\frac{3}{2}^-$ & & $-334$ & & 10629  \\
$\left[\eta_{\rm b}(2S) \otimes N\right]_P$ & & 23 & & $\frac{1}{2}^+$ or $\frac{3}{2}^+$ & & $-152$ & & 10786  \\
$\chi_{\rm b0}(1P) \otimes N$ & & 14 & & $\frac{1}{2}^+$ & & $-10$ & & 10788  \\
$\left[\Upsilon(2S) \otimes N\right]_P$ & & 23 & & $\frac{1}{2}^+$, $\frac{3}{2}^+$ or $\frac{5}{2}^+$ & & $-152$ & & 10810  \\
$\chi_{\rm b1}(1P) \otimes N$ & & 14 & & $\frac{1}{2}^+$ or $\frac{3}{2}^+$ & & $-10$ & & 10821  \\
$h_{\rm b}(1P) \otimes N$ & & 14 & & $\frac{1}{2}^+$ or $\frac{3}{2}^+$ & & $-10$ & & 10828 \\
$\chi_{\rm b2}(1P) \otimes N$ & & 14 & & $\frac{1}{2}^+$, $\frac{3}{2}^+$ or $\frac{5}{2}^+$ & & $-10$ & & 10841  \\
\hline
$\eta_{\rm b}(2S) \otimes N$ & & 33 & & $\frac{1}{2}^-$ & & $-545$ & & 10393  \\
 & &  & &  & & $-30$ & & 10909  \\
$\Upsilon(2S) \otimes N$ & & 33 & & $\frac{1}{2}^-$ or $\frac{3}{2}^-$ & & $-545$ & & 10418  \\
 & &  & &  & & $-30$ & & 10933  \\
$\left[\eta_{\rm b}(2S) \otimes N\right]_P$ & & 33 & & $\frac{1}{2}^+$ or $\frac{3}{2}^+$ & & $-343$ & & 10595  \\
$\left[\Upsilon(2S) \otimes N\right]_P$ & & 33 & & $\frac{1}{2}^+$, $\frac{3}{2}^+$ or $\frac{5}{2}^+$ & & $-343$ & & 10619  \\
$\chi_{\rm b0}(1P) \otimes N$ & & 21 & & $\frac{1}{2}^+$ & & $-116$ & & 10682  \\
$\chi_{\rm b1}(1P) \otimes N$ & & 21 & & $\frac{1}{2}^+$ or $\frac{3}{2}^+$ & & $-116$ & & 10715  \\
$h_{\rm b}(1P) \otimes N$ & & 21 & & $\frac{1}{2}^+$ or $\frac{3}{2}^+$ & & $-116$ & & 10722 \\
$\chi_{\rm b2}(1P) \otimes N$ & & 21 & & $\frac{1}{2}^+$, $\frac{3}{2}^+$ or $\frac{5}{2}^+$ & & $-116$ & & 10735  \\
\hline
\hline
\end{tabular}
\caption{Baryo-bottomonium model predictions (fourth and fifth columns), calculated by solving the Schr\"odinger equation (\ref{eqn:Hhc}) with the chromo-electric polarizabilities $\alpha_{\Upsilon \Upsilon}(2S)$ or $\alpha_{\Upsilon \Upsilon}(1P)$. In some cases, e.g. $\Upsilon(2S) \otimes N$, the baryo-bottomonium potential well is deep enough to give rise to a ground state plus its excited state.}
\label{tab:baryo-bottomonium-spectrum}
\end{table*}
%%%%%%%%%%%%%%%%%%%%%%%%%%

In order to show that the binding energies of the $b \bar b - qqq$ system are strongly dependent on the values of the chromo-electric polarizabilities, we present two sets of results for the spectrum of this system. The results are listed on the fourth column of Table \ref{tab:baryo-bottomonium-spectrum}.
In the first case, we use $\alpha_{\Upsilon \Upsilon}(2S)$ and $\alpha_{\Upsilon \Upsilon}(1P)$ from Eqs. (\ref{eqn:alpha-2S-min}) and (\ref{eqn:alpha-1P-min}), respectively; in the second case, the values of $\alpha_{\Upsilon \Upsilon}(2S)$ and $\alpha_{\Upsilon \Upsilon}(1P)$ are given by Eqs. (\ref{eqn:alpha-2S-max}) and (\ref{eqn:alpha-1P-max}), respectively.
The two sets of $\alpha_{\Upsilon \Upsilon}$'s values are calculated by using two different definitions of the Bohr radius of bottomonium; see Eqs. (\ref{eqn:a0-Bohr}) and (\ref{eqn:a0-Bohr2}).

Our results for the baryo-bottomonium pentaquarks span a wide energy interval, $10.4 - 10.9$ GeV.
The presence of a heavier (nonrelativistic) $Q \bar Q$ pair is expected to make the system more stable: this is why the hidden-bottom pentaquarks are more tightly bounded than their hidden-charm counterparts. See Tables \ref{tab:baryo-charmonium-spectrum} and \ref{tab:baryo-bottomonium-spectrum}. 
Our conjecture agrees with an unitary coupled-channel model \cite{Shen:2017ayv}, and with a molecular model \cite{Yamaguchi:2016ote} approach. 
For this reason, after the recent observation of hidden-charm $P_{\rm c}(4380)$ and $P_{\rm c}(4450)$ pentaquarks, we suggest to experimentalists that looking for pentaquark states in the hidden-bottom sector may be essential on the analysis of new bound states.
Moreover, in some cases, the baryo-bottomonium potential well is deep enough to give rise to a ground state plus its excited state, as can be observed in the case of $\eta_{\rm b}(2S) \otimes N$ and $\Upsilon(2S) \otimes N$ from the lower half of Table~\ref{tab:baryo-bottomonium-spectrum}, though the binding energy of the excited state is just a few tens of MeV. The emergence of these excitations is a consequence of the value of the bottomonium chromo-electric polarizability. If the value of $\alpha_{\Upsilon \Upsilon}(2S)$ is decreased from 33 to 23 GeV$^{-3}$, these excitations disappear.

Once the values of the $2S$ and $1P$ bottomonium chromo-electric polarizabilities are fitted to available experimental data, it will be interesting to discuss the possible emergence of deeply bound baryo-bottomonium pentaquarks by using more realistic values of $\alpha_{\Upsilon \Upsilon}(n\ell)$.

\section{Conclusion}
We adopted the baryo-charmonium model to discuss the possible emergence of $\eta_{\rm c}$- and $J/\psi$-$N^*$, $\eta_{\rm c}(2S)$-, $\psi(2S)$- and $\chi_{\rm c}(1P)$-$N$ bound states, where $N$ is the nucleon and $N^*$ a nucleon resonance.
The energies of baryo-charmonia were computed by solving the Schr\"odinger equation for the baryo-charmonium potential \cite{Dubynskiy:2008mq,Ferretti:2018kzy}, which was approximated as a finite well whose width and size could be expressed as a function of the $N$ ($N^*$) radius and the charmonium chromo-electric polarizability, $\alpha_{\psi \psi}$.
The baryo-charmonium masses and quantum numbers were compared with the existing experimental data, so that we could interpret the recently discovered $P_{\rm c}(4380)$ and $P_{\rm c}(4450)$ pentaquarks as $\psi(2S) \otimes N$ and $\chi_{\rm c2}(1P) \otimes N$ baryo-charmonia, respectively.

\begin{figure}[htbp]
\centering
\includegraphics[width=8cm]{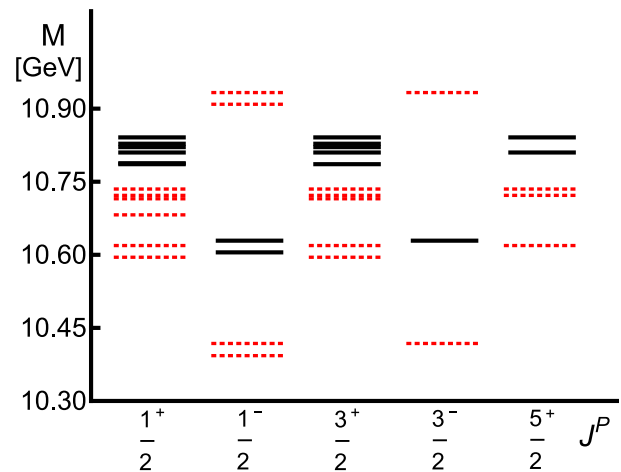}
\caption{Baryo-bottomonium spectrum of $\eta_{\rm b}(2S)$-, $\Upsilon(2S)$ and $\chi_{\rm b}(1P)$-$N$ bound states, calculated with: 1) $\alpha_{\Upsilon \Upsilon}^{\rm scatt}(2S) = 33$ GeV$^{-3}$ and $\alpha_{\Upsilon \Upsilon}^{\rm scatt}(1P) = 21$ GeV$^{-3}$ (dotted lines); $\alpha_{\Upsilon \Upsilon}^{\rm scatt}(2S) = 23$ GeV$^{-3}$ and $\alpha_{\Upsilon \Upsilon}^{\rm scatt}(1P) = 14$ GeV$^{-3}$ (black lines).}
\label{fig:bb-spectrum}
\end{figure}

We also provided results for bottomonium-nucleon bound states, which we suggest the experimentalists to look for in the $10.4-10.9$ GeV energy region.
The beauty partners of the LHCb pentaquarks, $P_b$, were found to be more deeply bound. In some cases, the potential well describing the interaction between the bottomonium core and the baryonic matter was found to be deep enough to give rise to a ground- plus excited state. 
For this reason, we believe that it is more probable to detect hidden-bottom pentaquarks than their hidden-charm counterparts. 
On the other hand, if the quarkonium-nucleon binding energy becomes too large, the baryo-quarkonium picture could be broken down. 
As a consequence, the quarkonium and baryon components may overlap, and a compact five-quark state could be realized rather than a baryon-meson bound state. In the case of a compact state, the interaction picture may also be different; for instance, the one gluon exchange would not be negligible. This possibility is worth to be investigated in more details.

Coupled-channel effects are not considered in the present study. We believe that the possibility of introducing them or writing the baryo-quarkonium potential in a slightly different way will be worth investigating once the value of the chromo-electric polarizability is adjusted to the experimental data.

In conclusion, a possible way to disentangle the internal structure of the baryo-quarkonium pentaquarks is to study their strong decay patterns in the baryo-quarkonium picture. This will be the argument of a subsequent paper.

\section*{Acknowledgments}
This work is supported in part by the Sino-German Collaborative Research Center ``Symmetries and the Emergence of Structure in QCD" (NSFC Grant No. 11621131001, DFG Grant No. TRR110), by the NSFC (Grant No. 11747601), and by the CAS-TWAS President's Fellowship for International Ph.D. Students.

\end{document}